\documentclass[12pt]{article}
\usepackage{epsfig,latexsym,amssymb}

\def\t{\tau}
\def\th{\theta}
\def\ra{\rightarrow}

\newcommand{\be}{\begin{equation}}
\newcommand{\ee}{\end{equation}}
\newcommand{\bea}{\begin{eqnarray}}
\newcommand{\eea}{\end{eqnarray}}

\def\ul{\underline}

\def\a{\alpha}
\def\b{\beta}

\def\d{\delta}

\def\s{\sigma}

\def \t {\tau}

\def\Tr{\rm Tr}
\def\pa{\partial}

\def\none{${\cal N}=1$ }

\begin{document}

\begin{titlepage}
\hfill IC/2004/12
\par\hfill CPHT-RR-009-0304
\par\hfill IFUM-787-FT\\
\vskip .1in \hfill hep-th/0403261

\hfill

\vspace{20pt}

\begin{center}
{\Large \textbf{Semiclassical spinning strings}}
{\Large \textbf{and confining gauge theories}}
\end{center}

\vspace{6pt}

\begin{center}
\textsl{F. Bigazzi ${}^{a}$, A. L. Cotrone ${}^{b,c}$ and L. Martucci${}^{d}$} \vspace{20pt}

\textit{a) The Abdus Salam International Centre for Theoretical Physics, Strada Costiera, 11; I-34014 Trieste, Italy.}\\
\textit{b) Centre de Physique Th\'eorique, \`Ecole Polytechnique, 48 Route de Saclay; F-91128 Palaiseau Cedex, France.}\\
\textit{c) INFN, Piazza dei Caprettari, 70; I-00186  Roma, Italy.}\\
\textit{d) Dipartimento di Fisica, Universit\`{a} di Milano, Via Celoria, 16; I-20133 Milano, Italy.}
\end{center}

\vspace{12pt}

\begin{center}
\textbf{Abstract }
\end{center}

\vspace{4pt} {\small \noindent
We study multi-charged rotating string states on Type IIB regular backgrounds dual to confining $SU(N)$ gauge theories with (softly broken) ${\cal N}=1$ supersymmetry, in the infra red regime.
After exhibiting the classical energy/charge relations for the folded and circular two-charge strings, we compute in the latter case the one loop sigma-model quantum correction.
The classical relation has an expansion in positive powers of the analogous of the BMN effective coupling, while the quantum corrections are non perturbative in nature and are not subleading in the limit of infinite charge. We comment about the dual field theory multi-charged hadrons and the implications of our computation for the $AdS/{\cal N}=4$ duality.
}
\vfill
\vskip6mm
{\small
\noindent
bigazzif@ictp.trieste.it; Cotrone@cpht.polytechnique.fr; Luca.Martucci@mi.infn.it}
\end{titlepage}

 \setcounter{page}{1}
 \section{Introduction}
The string/gauge theory correspondence \cite{maldaetal} was recently put at work beyond the supergravity approximation of string theory \cite{BMN,GKP}. A crucial role in the game was played by special sets of dual string/gauge theory states parameterized by large quantum numbers. In the master example
of the $AdS/CFT$ correspondence, a limit was discovered where
the energy of large angular momentum closed string states (rotating
along a great circle of $S^5$) reproduces the 
conformal dimensions of the corresponding gauge theory operators. The
string states in the limit are semiclassical solitons, whose physics
is effectively described by string theory on the so called Penrose
limit \cite{penrose} of the $AdS_5 \times S^5$ original background. This results in a
parallel plane (pp) wave background \cite{blau} on which string theory can be exactly solved \cite{metse}. This stimulated the study of more general (multi)spinning
string states in $AdS_5 \times S^5$ (see the review
\cite{tseyrew} and references therein\footnote{More recent solutions, not included in the references of \cite{tseyrew}, can be found in \cite{tutti}.})
whose simple classical description often provides 
non trivial predictions on the dual quantum gauge
theory.

A lot of work is still required in order to extend the
above techniques beyond the master conformal example. 
The present paper is devoted to the study of multi-spinning string solutions on the regular IIB string backgrounds conjectured to
be dual to confining, four dimensional, $SU(N)$ gauge theories\footnote{Other classical solutions in confining backgrounds 
were studied in \cite{crucco,pons,Alishahiha:2004vi}.}.
We will concentrate on string solutions relevant for the study of the low
energy gauge theory dynamics. In the regime covered by the
semiclassical string approximation, the gauge theories we will examine
will be coupled to towers of massive (Kaluza-Klein) adjoint fields generically charged under some $U(1)$ symmetry group. The semiclassical closed string solutions describe large (multi)charged massive hadronic states made of these particles.

For all the confining cases present in literature, when the gauge theories are strongly
coupled the dual backgrounds asymptotically reach a relatively simple,
universal form in the IR. Very roughly, it is a flat $d+1$ dimensional Minkowski
space-time, times a flat $q$ dimensional space, times a transverse
$9-d-q$ sphere (usually those are not really products spaces, but there is some nontrivial mixing between the various factors). The isometries of the sphere correspond to global
symmetries for the adjoint Kaluza-Klein matter fields. It is not
complicated to look for string solutions on those backgrounds.  They
should describe gauge theory objects visible in the low energy
limit. In the papers \cite{pandostras,abc,sonne,univers} {\it point-like} strings sitting at the origin of
the $q$ dimensional space\footnote{This corresponds to the
  low energy region of the dual gauge theory.} and rotating along a great
circle of the transverse sphere, were considered. Just as in the
$AdS_5\times S^5$ case, the semiclassical approximation here is effectively described 
by the string quantization on a generalized
Penrose limit of the corresponding original backgrounds. From the
gauge theory point of view, the string Hamiltonian describes the non
relativistic motion (in a $d$-dimensional space) and excitations of
stringy shaped {\it hadrons}. Their constituents are the adjoint
Kaluza-Klein massive fields. String theory provides a non trivial
prediction on (the relations between) masses and charges of these
hadronic states.

In the present paper we study more general circular and
folded\footnote{The classical solution for the single charge folded string was presented in \cite{pons}.} closed string solutions with multiple spins along the transverse
sphere in our preferred example (for its simplicity) of the Maldacena-N\`u\~nez (MN) \cite{mn} solution and its non supersymmetric version (bMN) \cite{gubser,bMN}. The former is conjectured to be dual to an ${\cal N}=1$ $SU(N)$ gauge theory in four dimensions. The latter is dual to the same gauge theory in which supersymmetry is broken by a gluino mass term. In both cases the transverse sphere is an $S^3$ and so our general solutions will have two spins along it. Analogous solutions for the (softly broken) Klebanov-Strassler\footnote{For a review of the (b)MN and KS models see for example \cite{noirev}.} (KS \cite{ks} and bKS \cite{borgubs,bKS}) case are exactly the same as the (b)MN ones (the relevant background in the far IR is the same) and so we do not present the (b)KS results explicitly.
As expected from the symmetries of the problem, the integrable structures playing a crucial role in the example of the AdS/CFT
correspondence \cite{frolov5,russo,tseyrew} are manifest in the confining context
too.
In fact, the trivial but crucial observation that the deep IR geometry of confining backgrounds is of the factorized ``flat space $\times$ a sphere'' form, makes it evident that the {\it classical} solutions will be the same as the ones studied in the $AdS/{\cal N}=4$ context, with only a trivial difference in some coefficients. 
Thus, the classical energy/charge relations for these configurations admit an expansion in positive powers of the effective coupling\footnote{Here $\lambda=e^{\Phi_0}N$, $\Phi_0$ being the value of the dilaton at the origin, and $J$ is the sum of the charges. Note that the power of $\lambda$ is different from the $AdS$ case because of the presence of the three-form instead of the five-form field strength \cite{pandostras}.} $\lambda^2/J^2$. An expansion of this kind is usually called {\it regular}. It is obtained by taking the large $\lambda,J$ limit, with $\lambda^2/J^2$ fixed, of the classical energy of the string states. The leading term in the limit gives a linear relation between the energy and the total charge of the dual hadrons. We will speculate on a possible interpretation of the regular corrections to this relation as due to the collective binding energy of the hadrons. 
 
The main part of the paper is devoted to the calculation of the quadratic fluctuations around the circular solutions on the bMN background. We will study their quantization and examine the one loop contribution to the zero-point energy.
The latter turns out to be non-vanishing for generic values of the supersymmetry breaking parameter and in particular in the supersymmetric case. As a consequence, the energy/charge relations are corrected at the quantum level for these configurations, as it happens in the special case of the point-like, single-charge string \cite{abc}. Moreover the one loop corrections are different in nature with respect to the classical ones. In fact, the leading order quantum corrections to the $E\approx J$ relation scale as $1/\lambda$ and thus are not ``regular'' as the latter. 
Also it results that, contrarily to what happens in the $AdS_5\times S^5$ context \cite{frolov2,frolov3,tseyrew}, these quantum corrections (which are anyway subleading with respect to the classical terms because $\lambda\gg 1$), are not subleading in the limit of infinite total charge $J\to\infty$.
As a consequence, the classical result cannot be extrapolated to small values of $\lambda$. 
We will speculate on the possible interpretation of the one-loop string sigma model contributions as non perturbative (in $\lambda$) corrections to the masses of the single constituents. 

Even if our energy/charge relations cannot be compared with perturbative field theory calculations as in the $AdS/CFT$ case, they provide new non-trivial predictions for the mass/charge (Regge-like) relations for the dual hadrons in the strong coupling regime. 
As pointed out in \cite{univers}, lattice simulations could in principle provide an ``experimental'' test for the validity of our results for the string/confining gauge theory correspondence in this sector.

Moreover, our calculation turns out to be relevant for a better understanding of the $AdS/CFT$ case too.
While in the literature there are plenty of classical solutions, the explicit calculations of the quantum corrections have been done in very few examples \cite{frolov1,frolov3}.
We provide one such computation in a context which is slightly different from  $AdS$.
This allows us to cast some conclusions on the nature of the one loop correction and its implication for the regularity of the expansion of the energy/charge relation in general backgrounds. 

In section \ref{sezmnback} we present the (b)MN background in the IR in a form which is useful for the quantization.
Then, in section \ref{sezclass}, we review the connection of the spinning strings with integrable systems (sec. \ref{sezinteg}) and show the folded (sec. \ref{sezino}) and circular (sec. \ref{sezcirc}) string solutions, together with the classical energy/charge relations in both cases.
The computation of the one loop correction to this relation for the circular string case is presented in section \ref{sezsemi}, while in section \ref{sezsusy} we give a simple argument for the absence of supersymmetry in our solutions.
We discuss the dual multi-charged hadrons in section \ref{sezmulti} and conclude with some comments about the implications of our results in section \ref{sezconc}.
We also include an appendix with the Green-Schwarz superstring on general backgrounds (to second order in the fermions) and one with a brief discussion of the UV finiteness and stability of the circular solution.


\section{The (b)MN background}\label{sezmnback}

Let us start by briefly reviewing the basic aspects of the MN solution \cite{mn} and its softly broken generalization \cite{gubser,bMN}.   
In the string frame, the MN solution in global coordinates is given by  
\bea
\label{MNsol}
ds^2_{str} &= & e^{\Phi } \Big\{ dx_\mu dx^\mu
+ \alpha' N[ d \rho^2 + e^{ 2 g(\rho)}
(d\theta_1^2+ \sin^2\theta_1 d\phi_1^2)+
{1 \over 4 } \sum_a (w^a - A^a)^2 ] \Big\}\ ,
\\
F^{(3)}& =& \alpha'N \Big[  -{1\over 4} (w^1 -A^1)\wedge (w^2 - A^2) \wedge ( w^3-A^3)  + { 1 \over 4}
\sum_a F^a \wedge (w^a -A^a) \Big]\ ,
\eea
with a $\rho$-dependent dilaton whose value at the origin is a continuum parameter $\Phi_0$ and gauge field $A$ given by
\be
\label{Afield}
A={1\over 2} \left[ \sigma^1 a(\rho) d \th_1
+ \sigma^2 a(\rho) \sin\th_1 d\phi_1 +
\sigma^3 \cos\th_1 d \phi_1 \right]\ .\ee
The one-forms $w^a$ are defined by 
\bea\label{unoforme}
 { i \over 2} w^a \sigma^a &  = &dg g^{-1}\ , \qquad  \qquad\qquad  \qquad  \qquad \quad g = e^{ \frac i2 \psi \sigma_3} e^{ \frac i2 \th_2 \sigma_1 }
e^{\frac i2\phi_2 \sigma_3}\ ,          \nonumber \\
w^1 + i w^2 & = &e^{ - i \psi } ( d \th_2 + i \sin \th_2 d \phi_2)  ~,~~~~~~~~~~
w^3 = d \psi +\cos \th_2  d\phi_2\ .
\eea
The full explicit form of the functions $g(\rho),\, \Phi(\rho),\, a(\rho)\,$ in the supersymmetric case is 
\begin{eqnarray}  \label{defins}
a(\rho) &=& \frac{2\rho}{\sinh{2\rho}}\ , \nonumber\\    
e^{2g}(\rho) &=& \rho \coth{2\rho}-\frac{\rho^2}{\sinh^2{2\rho}}-\frac{1}{4}\ ,\\   
e^{2\Phi}(\rho)& =&g_s^2\frac{\sinh{2\rho}}{2e^{g(\rho)}}\ ,\nonumber    
\end{eqnarray}
where it has to be noticed that what we define as the effective string coupling is really the arbitrary parameter $g_s=e^{\Phi_0}$. 

In the softly broken version \cite{gubser,bMN} we know in analytic form only the asymptotic behavior.
In the IR it reads
\begin{eqnarray}  
a(\rho) &=& 1 - b \rho^2 + ...\ , \nonumber\\    
e^g(\rho) &=&\rho - ( {b^2\over 4} +{1 \over 9}) \rho^3 + ...\ ,\\   
\Phi(\rho)& =& \Phi_0 + ({b^2\over 4} + {1\over 3}) \rho^2 + ...\ .\nonumber    
\end{eqnarray}
The correspondence with the dual field theory is clear only for $b\in (0,2/3]$, the value $b=2/3$ corresponding to the supersymmetric solutions above\footnote{It will be interesting, from the string point of view, to allow $b$ to extend in the full range $[0,2]$ giving regular solutions \cite{gubser}.}.
In order to explore the IR of the above background it is convenient to perform a gauge transformation on $A$ such that it actually goes to zero when $\rho\ra0$. This can be done since $A$, also in the softly broken case, is a pure gauge in the extreme IR \cite{mn}.
We use the gauge transformation \cite{bertmerl} $A\rightarrow h^{-1}A h+ih^{-1}dh$ with $h=e^{i\sigma^1\th_1/2}e^{i\sigma^3\phi_1/2}$. The resulting field has the following expression in the $\rho\ra0$ approximation
\begin{eqnarray}
\label{AIR2}
A &=& \Big(-{b\over 2}\rho^2 + {\mathcal{O}}(\rho^4)\Big)
\Big[\sigma^1(\cos\phi_1\,d\theta_1- \cos\th_1
\sin\th_1 \sin\phi_1\,d\phi_1 ) \nonumber\\
&&\; + \,\sigma^2(\sin\phi_1\,d\theta_1 + \cos\theta_1
\sin\theta_1 \cos\phi_1\,d\phi_1 ) +
\,\sigma^3(\sin^2\theta_1\,d\phi_1)\Big]\ .
\end{eqnarray}
In the same regime, we  have that
\bea
e^{\Phi}=g_s\left[ 1+({b^2\over 4} + {1\over 3}) \rho^2 +{\cal O}(\rho^4)\right]\ .
\eea
Since the coordinates $\rho,\ \theta_1,\ \phi_1$ are clearly degenerate at $\rho=0$, it is convenient to use the almost Euclidean  coordinates $y^a$, with $a=1,2,3$, related to  $\rho,\theta_1,\phi_1$ as if these were ordinary spherical coordinates on $\mathbb{R}^3$ and the $y^a$'s the flat ones. Then, in the gauge defined above, we have that
\bea
A^1&=&-b(y^3dy^1-y^1dy^3)\ ,\cr
A^2&=&+b(y^2dy^3-y^3dy^2)\ ,\cr
A^3&=&-b(y^1dy^2-y^2dy^1)\ .
\eea     
Furthermore, the only dimensionful coordinates are the $x^\mu$ that parameterize the four flat  directions. Then, in order to study the string world-sheet theory on this background, it is more comfortable to rescale these flat coordinates in new dimensionless coordinates $X^\mu=m_0\,x^\mu$, where $m_0=1/\sqrt{\alpha^\prime N}$ is the mass scale of the glueball and KK fields of the dual SYM theory\footnote{We will think to the KK and glueball scale $m_0$ as the natural scale of the theory. We will work with dimensionless quantities and  the right dimensionality will be given by reintroducing $m_0$ opportunely.}. In this way, the bMN metric (up to quadratic order in $y^a$) takes the form             
\bea
ds^2 &=&\frac{g_s}{m_0^2}\Big\{\big[1+({b^2\over 4} + {1\over 3}) y^ay^a\big] dX^\mu dX_\mu +dy^ady^a +\frac14\big[1+({b^2\over 4} + {1\over 3}) y^ay^a\big]w^aw^a
 +\cr 
&&+\frac b2\epsilon_{ab2}y^ady^bw^1-\frac b2\epsilon_{ab1}y^ady^bw^2+\frac b2\epsilon_{ab3}y^ady^bw^3\Big\}\ .
\eea
It is the product of flat 4d Minkowski space-time with an $S^3$ fibration in a six dimensional manifold.
At the point $y^a=0$ it degenerates into ${\mathbb R}^{3,1}\times S^3$.

Let us now observe that in these coordinates the bMN solution has the killing directions $\partial_{\phi_2}$ and $\pa_\psi-\pa_{\phi_1}$.
In order to make these symmetries manifest, it is convenient to go to a rotated frame in the $(y^1,y^2)$ plane, i.e. we introduce new coordinates $Y^a$ defined by
\bea
y^1+iy^2=e^{-i\psi}\big(Y^1+iY^2\big)\ ,\quad\quad y^3=Y^3\ .
\eea
It is also  useful to introduce the following coordinates on $S^3$
\bea\label{s32}
\chi=\frac{\theta_2}{2}\ ,\quad\quad \phi_+=\frac12 (\psi+\phi_2)\ ,\quad\quad \phi_-=\frac12 (\psi-\phi_2)\ .
\eea
Then the metric expanded up to the second order in $Y^a$ assumes the following form
\bea\label{metrnew2}
ds^2&=&\frac{g_s}{m_0^2}\Big\{ \big[1+({b^2\over 4} + {1\over 3}) Y^a Y^a\big] dX^\mu dX_\mu +dY^adY^a +\cr
&&+\big[1+({b^2\over 4} + {1\over 3}) Y^aY^a\big](d\chi^2+\sin^2\chi d\phi_-^2+\cos^2\chi d\phi_+^2)+\cr
&&+[(Y^1)^2+(Y^2)^2][(1-\frac b2)(d\phi_++d\phi_-)^2-\frac b2 (\cos^2\chi-\sin^2\chi)d(\phi_++\phi_-)d(\phi_+-\phi_-)]+\cr 
&&+ (Y^1dY^2-Y^2dY^1)[(\frac b2-2)(d\phi_++d\phi_-)+\frac b2 (\cos^2\chi-\sin^2\chi)d(\phi_+-\phi_-)]+\cr 
&&+ \frac b2[-2\sin\chi\cos\chi Y^1Y^3d(\phi_++\phi_-)d(\phi_+-\phi_-)+ 2Y^2 Y^3d\chi d(\phi_++\phi_-)+\cr
&&+ 2(Y^3dY^1-Y^1dY^3)d\chi -2\sin\chi\cos\chi(Y^2dY^3-Y^3dY^2)d(\phi_+-\phi_-)]\Big\}\ .
\eea
In the following we will need the RR 3-form $F^{(3)}$ only at leading order. In the new coordinates it is given by
\bea\label{3str}
F^{(3)}&=&\frac1{4m_0^2}[-w^1\wedge w^2\wedge w^3 + \sum_a F^a\wedge w^a]\cr
       &=&\frac{1}{m_0^2}\big[2\sin\chi\cos\chi d\chi\wedge d\phi_- \wedge d\phi_+ -b\;dY^1\wedge dY^2 \wedge (\cos^2\chi d\phi_+ + \sin^2\chi d\phi_-)  +\cr
        && + b\; dY^1\wedge dY^3 \wedge d\chi +b\sin\chi\cos\chi dY^2\wedge dY^3 \wedge (d\phi_+-d\phi_-)\big]\ .
\eea 
As can be seen directly from (\ref{metrnew2}) and (\ref{3str}), $\pa_{\phi_+}$ and $\pa_{\phi_-}$ are the killing vectors of the internal part of our background. We will use the associated conserved angular momenta $J_+$ and $J_-$ to classify different classical string solutions and the semiclassical quantization of a subclass of them.

\section{Classical string solutions on the (b)MN background}\label{sezclass}

In this paper  we will study  a class of solutions describing  strings sitting at the point $Y^a=0$ and extending and spinning  along the $S^3$ directions of the transverse space. In particular, we are interested in string configurations having definite angular momenta $J_+$ and $J_-$ in the internal space. As we will discuss better in section \ref{sezmulti}, these solutions have a natural dual interpretation in terms of hadronic states made up of KK matter of the dual field theory and generalize the annulons discussed in \cite{pandostras,abc}. Those papers considered the Penrose limit of the background, equivalent to the quadratic expansion around the following point-like solution moving in the internal space 
\bea\label{collapsed}
X^0\equiv t=k\tau\ ,\quad\quad \phi_+=\nu\tau\ ,\quad\quad
\chi=\phi_-=0\ ,\quad \quad Y^a=X^a=0 \quad {\rm for} \ a=1,2,3\ .
\eea
The conformal gauge constraint 
implies that the solution must describe a null geodesic, ${\nu^2 } - k^2=0$.
The classical energy and charge of this solution are $E=\lambda k,\,J_+=\lambda\nu$.
Reintroducing the right dimensions using the fundamental scale $m_0$, we obtain the pp-wave relation $E=m_0 J_+$.

Let us observe  that $\lambda$ represents both the square of the ratio between the typical radius of curvature of the bMN solution and the fundamental string length, and the ratio between the YM string tension and the square of the KK mass scale $m_0^2$. Then the supergravity approximation is valid for $\lambda \gg 1$, i.e. in a regime opposite to the limit $\lambda\ll 1$ in which the KK states decouple from the pure ${\cal N}=1$ SYM. The dependence on the parameter $\lambda$ in the string world-sheet action can be easily isolated by using the dimensionless metric defined by the rescaling $ds^2\rightarrow (g_s/m_0^2)ds^2$. Then for example the Nambu-Goto (NG) bosonic string action (obviously, nothing changes in the Polyakov approach) becomes
\bea
S_B=-\frac{\lambda}{2\pi}\int d\tau\int_0^{2\pi}d\sigma \sqrt{-h}\ ,
\eea
where $h=\det(h_{\alpha\beta})$ and $h_{\a\b}$ is the induced world-sheet metric. 

If ${\cal X}^m$ denotes the general ten dimensional spacetime coordinate, we can study the dynamics around any given classical solution $\bar{\cal X}^m$ by performing an expansion ${\cal X}^m=\bar{\cal X}^m+\lambda^{-1/2}\delta {\cal X}^m$. In the supergravity approximation $\lambda\gg 1$ the leading contributions in the world-sheet action come from the action expanded up to the quadratic order in the fluctuating fields, i.e. from the one-loop approximation. Analogously to the case of the collapsed string in $AdS_5\times S^5$ \cite{GKP,frolov1}, performing such an expansion around the null geodesic (\ref{collapsed}) naturally produces the string theories on the pp-wave backgrounds studied in the \cite{pandostras,abc}. 

In the following sections we present some multispin solutions that generalize the collapsed string solution (\ref{collapsed}), computing the corresponding generalization of pp-wave relation $E=m_0 J_+$, while  the semiclassical analysis of a class of regular circular solutions, analogous to those considered in \cite{frolov2,frolov3}, will be performed in section \ref{sezsemi}.   

\subsection{Classical multispin solutions and integrable systems}\label{sezinteg}
As said, we will focus on some special class of solutions corresponding to string located at $Y^a=0$, where the bMN background degenerates into ${\mathbb R}^{3,1}\times S^3$. 
The latter is true in the (b)KS case too, so everything we will say in the present and following subsections applies to this solution as well.
Let us first consider the classical solutions on general grounds, reviewing the discussion presented in \cite{frolov5,russo,tseyrew} and adapting to our case some of the results presented there. To this end, it is sufficient to consider  the following effective bosonic action obtained from a truncation of the conformally gauge fixed Polyakov action  
\bea\label{redac0}
S&=& \frac{\lambda}{2\pi} \int{d\tau}\int_0^{2\pi}d\sigma\Big[ -\frac12\partial^\alpha X^\mu\partial_\alpha X_\mu  -\frac12
\partial^\alpha Z^M\partial_\alpha Z^M +\frac\Lambda2 (Z^MZ^M-1)\Big]\ .
\eea
The Euclidean coordinates $Z^M$, $M=1,\ldots,4$, constrained by $Z^MZ^M=1$ (hence the Lagrangian multiplier $\Lambda$ in the action above), describe the transverse $S^3$. 
Their relation to the $(\chi,\phi_-,\phi_+)$ coordinates can be given as
\bea\label{trancoo}
&&Z^1=\sin\chi\cos\phi_-\ ,\quad\quad Z^2=-\sin\chi\sin\phi_-\ ,\cr
&&Z^3=\cos\chi\cos\phi_+\ ,\quad\quad Z^4=\cos\chi\sin\phi_+\ .
\eea 
The action (\ref{redac0}) is to be supplemented by the conformal gauge constraints
\bea\label{concon}
&&\dot X^\mu\dot X_\mu +X^{\prime \mu}X^{\prime}_{\mu}+\dot Z^M\dot Z^M +Z^{\prime M}Z^{\prime M}=0\ ,\cr
&& \dot X^\mu X^\prime_\mu +\dot Z^M Z^{\prime M}=0\ ,\cr
&& Z^MZ^M=1\ .
\eea

We will look for rotating string solutions of the form 
\bea\label{ansatz0}
&&X^0\equiv t=k\tau\ ,\qquad\qquad X^a=0 \quad a=1,2,3\ ,\cr
&&Z^1+iZ^2 =r_1(\sigma)e^{iw_1\tau +i\alpha_1(\sigma)}=[z^1(\sigma)+iz^2(\sigma)]e^{iw_1\tau}\ ,\quad\cr  
&& Z^3+iZ^4 =r_2(\sigma)e^{iw_2\tau +i\alpha_2(\sigma)}=[z^3(\sigma)+iz^4(\sigma)]e^{iw_2\tau}\ .
\eea
From the condition $Z^MZ^M=1$ we must have
$r_1^2 + r_2^2=1$,
and the periodicity conditions are satisfied if
\bea\label{peco}
r_i(\sigma+2\pi)=r_i(\sigma)\ ,\quad\quad \alpha_i(\sigma +2\pi)=\alpha_i(\sigma)+2\pi m_i\ ,
\eea
where $m_i\in \mathbb{Z}$ play the role of winding numbers. 
The energy and the nonzero spins are 
\bea\label{cariche}
E=P^t=\lambda k\ ,\quad\quad J^{12}=\lambda w_1\int_0^{2\pi} \frac{d\sigma}{2\pi}r_1^2(\sigma)\ ,\quad\quad  
J^{34}=\lambda w_2\int_0^{2\pi} \frac{d\sigma}{2\pi}r_2^2(\sigma)\ ,
\eea
where the spins $J^{12}$ and $J^{34}$ satisfy the relation
\bea
\frac{J^{12}}{\lambda w_1}+\frac{J^{34}}{\lambda w_2}=1\ .
\eea
All the other charges vanish. It  is also useful to associate to any conserved charge $Q$ a rescaled charge ${\cal Q}$ defined as
\bea\label{defresc}
Q=\lambda{\cal Q}\ .
\eea 

Substituting the ansatz (\ref{ansatz0}) in the action (\ref{redac0}) one obtains the following ``effective'' one dimensional Lagrangian
\bea\label{acN}
L=\frac12 (z^{\prime M}z^{\prime M}- \omega_M^2 z^Mz^M)-\frac12 \Lambda(z^Mz^M-1)\ ,
\eea
where $\omega_1=\omega_2=w_1$ and $\omega_3=\omega_4=w_2$. This action corresponds to the well known $n=4$ Neumann integrable system. 
One can get more insights by writing the effective action (\ref{acN}) more explicitly in terms of the variables $r_i$ and $\alpha_i$
\bea\label{act3}
L=\frac12\sum_{i=1,2}\big( r_i^{\prime 2}+ r_i^2 \alpha_i^{\prime 2}-w_i^2r_i^2\big)-\frac12 \Lambda\big( \sum_{i=1,2}r_i^{2}-1\big)\ .
\eea  
This action implies immediately that $\alpha_i^{\prime }=v_i/r_i^2$, where $v_i$ are integrals of motion. We can directly use them by rewriting the effective action as
\bea\label{inteff}
L=\frac12\sum_{i=1,2}\big( r_i^{\prime 2}-w_i^2r_i^2-\frac{v_i^2}{r_i^2}\big)-\frac12 \Lambda\big( \sum_{i=1,2}r_i^{2}-1\big)\ .
\eea   
The conformal gauge constraints (\ref{concon}) imply that 
\bea\label{coco}
k^2=\sum_{i=1,2}( r_i^{\prime 2}+w_i^2r_i^2+\frac{v_i^2}{r_i^2}\big), \qquad \quad \sum_{i=1,2}w_iv_i=0\ .
\eea
From the second of these relations we see that the two integrals of motion $v_i$ are not independent. 

As a last step, let us note that we can generalize any given solution $Z^M\equiv\bar Z^M(\tau,\sigma)$ of the form (\ref{ansatz0}), considering a more general $X^\mu(\tau,\sigma)$. 
This class of solutions, which includes the multi-charged generalization of the one describing the usual (i.e. mass squared vs. spin) Regge trajectory \cite{psv}, is classically given by the standard flat-space mode expansion for the coordinates $X^\mu(\tau,\sigma)$.
The only difference with respect to the flat space solution is in the conformal constraint, giving the modified on-shell condition
\be
m^2=-P^\mu P_\mu 
=\lambda^2\big[k^2+4\sum_{n\geq 1}(\alpha^\mu_n \alpha_{-n\mu}+\tilde \alpha^\mu_n \tilde \alpha_{-n\mu} )\big]\ ,
\ee
the $k$ factor being given by the internal part of the solution, (\ref{ansatz0}).
It would be very interesting to perform a semiclassical analysis of this generalized solution along the lines of section \ref{sezsemi}, but we postpone this problem to the future.

In the following subsections we will consider two sub-cases of possible solutions of the integrable system (\ref{inteff}). The first is obtained by posing $v_i=0$ and gives the generalization of the point-like solution to multi-spin folded strings. The second family of solutions, presented in subsection \ref{sezcirc}, is constituted by circular solutions analogous to the multi-spin solutions of \cite{frolov2}. Due to the peculiar regularity of these solutions, we choose them as natural candidates to explore the one-loop extension of the classical results. This point will be discussed in section \ref{sezsemi}.


\subsection{Folded solutions}\label{sezino}

Let us start by considering the case of $v_i=0$. Then, the effective action (\ref{act3}) reduces to an $n=2$ Neumann model,
\bea\label{inteff2}
L=\frac12\sum_{i=1,2}\big( r_i^{\prime 2}-w_i^2r_i^2\big)-\frac12 \Lambda\big( \sum_{i=1,2}r_i^{2}-1\big)\ .
\eea   
Instead of giving the general solution of this integrable system, we focus on the simplest one. 
The condition $v_i=0$ imply that the angles $\alpha_i$ are constant. 
We then fix them to zero and from the relations (\ref{trancoo}) we see that we are looking for solutions of the form
\be
\theta_2(\sigma)=\theta_2(\sigma+2\pi)\ , \qquad t=k\tau\ , \qquad \phi_+=\nu\tau\ ,\qquad \phi_-=\omega\tau\ ,
\ee
where $\nu=w_2$, $\omega=-w_1$ and  we have used $\theta_2=2\chi$ instead of $\chi$.
The effective action 
\bea
L=(\theta_2^\prime)^2+2(\omega^2-\nu^2)\cos{\theta_2}\ ,
\eea
is then equivalent to the classical action for a  pendulum in a constant gravitational field, where $\sigma$ plays the role of the time and $V(\theta)\equiv-2(\omega^2-\nu^2)\cos{\theta_2}$ is the  potential energy. This system can be easily integrated using the  energy integral of motion $e$, which is related to $k,\ \omega$ and $\nu$ by the conformal constraint  
\be\label{constraint2}
e\equiv\theta_2'^2-2(\omega^2-\nu^2)\cos{\theta_2}=4 k^2  - 2(\omega^2+\nu^2)\ .
\ee

The charges of these solutions are given by\footnote{In this paper we choose to work with the following sign conventions: $w_1,\, w_2,\, m_1\geq 0$ and $m_2\leq 0$. As a consequence, we will have $J_+\geq 0$ and $J_-\leq 0$.}
\bea\label{chs}
E &=& \lambda k\ ,\cr
J_+ &=& \frac{\lambda\nu}{4\pi}\int_0^{2\pi}d\sigma(1+\cos\theta_2(\sigma))\ ,\cr
J_- &=& \frac{\lambda\omega}{4\pi}\int_0^{2\pi}d\sigma(1-\cos\theta_2(\sigma))\ .
\eea
From these expressions we can easily recover the general relation
\be
E=\frac{k}{\nu}J_+ +\frac{k}{\omega}J_-\ .
\ee
We can restrict ourself to the case $\omega^2 \geq\nu^2$ without loosing in generality. In fact the case $\nu^2\geq\omega^2$ can be obtained from the previous one by means of the simple redefinitions
\be\label{ccoord}
\theta_2\rightarrow \theta_2+\pi\ ,\qquad \phi_+\leftrightarrow \phi_-\ ,
\ee  
that correspond to a $\pi/2$ rotation in the planes (3,1) and (2,4) of the ${\mathbb R}^4$ parameterized by the $Z^M$.

First of all, let us notice that in the case $\omega^2=\nu^2$ (absence of gravity), we have the solution
\be
\theta_2 = \pm 2\sigma\sqrt{k^2-\omega^2}+{\rm const}.\ ,
\ee
corresponding to closed circular strings. Taking  $k^2 =\omega^2$ we have collapsed strings, for which $E=J_+ - J_-\equiv J$. On the other hand the periodicity condition imposes as other possible choice only $k^2=1+\omega^2$ corresponding to circular strings with energy
\be\label{circularspecial}
E=\sqrt{m^2\lambda^2+J^2}\ , 
\ee 
being $m$ the number of windings.
Considering now the case $\omega^2>\nu^2$, we must have $e\geq V_{min}$, i.e. $k^2\geq \nu^2$, and we can consider three kinds of  different solutions:  
\begin{itemize}
\item 
$e=V_{min}$ i.e. $ k^2=\nu^2$. From (\ref{constraint2}) we obtain that the only admissible solution is given by $\theta_2(\sigma)\equiv 0$. These kind of solutions correspond to collapsed point-like strings (the pendulum does not oscillate).
\item
$V_{min}< e < V_{max}$ i.e.  $\nu^2<k^2<\omega^2$. The string is folded and arrives at a maximal angle $\theta_2^{max}<\pi$ (the pendulum does oscillate).  
\item
$e = V_{max}$ i.e.  $k^2=\omega^2$. The folded string is at its extreme limit (the pendulum is in the unstable vertical position). 
\item
$e > V_{max}$ i.e. $k^2>\omega^2$. The string is circular (the pendulum turns around with never vanishing angular velocity). 
\end{itemize}
Since the point-like solution can be considered as a limiting case of the folded one and the circular strings are unstable and will decay into the folded ones by ``slipping off the side'' \cite{GKP}, we focus our attention on the folded string case where $\nu^2<k^2<\omega^2$. 
The maximal angle $\theta_2^{max}<\pi$ is given by 
\be
\cos \theta_2^{max}=1-\frac{2(k^2-\nu^2)}{\omega^2-\nu^2}\ .
\ee
From (\ref{constraint2}) one can write the periodicity condition $\theta_2^{max}(\sigma)=\theta_2^{max}(\sigma+2\pi)$ in the form
\be\label{percond}
\frac\pi2 \sqrt{2(\omega^2-\nu^2)}=\int_0^{\theta_2^{max}}{\frac{d\theta_2}{\sqrt{\cos\theta_2-\cos\theta_2^{max}}}}\ .
\ee
From (\ref{chs}), (\ref{constraint2}) and (\ref{percond}) we can write the angular momenta in the following form
\bea\label{chs2}
J_+ &=& \frac{\lambda\nu}{\pi\sqrt{2(\omega^2-\nu^2)}}\int_0^{\theta_2^{max}}d\theta_2\frac{1+\cos\theta_2}{\sqrt{\cos\theta_2-\cos\theta_2^{max}}}\ ,\cr
J_- &=& \frac{\lambda\omega}{\pi\sqrt{2(\omega^2-\nu^2)}}\int_0^{\theta_2^{max}}d\theta_2\frac{1-\cos\theta_2}{\sqrt{\cos\theta_2-\cos\theta_2^{max}}}\ .
\eea
If we introduce the new parameter
\bea\label{dfn}
\eta=\frac12(1-\cos\theta_2^{max})=\frac{k^2-\nu^2}{\omega^2-\nu^2}\ ,\qquad 0<\eta<1\ ,
\eea
we can write (\ref{percond}) and (\ref{chs2}) in terms of hypergeometric functions
\bea\label{pps}
&&\sqrt{2(\omega^2-\nu^2)}=\sqrt{2}F(\frac12,\frac12,1;\eta)\ ,\cr
&&\cr
&&J_+ = {\lambda\nu}\big[ 1-\eta + \frac{\eta}{\sqrt{4(\omega^2-\nu^2)}} F(\frac12,\frac12,2;\eta)   \big]\ ,\cr
&&J_- = \frac{\lambda\omega}{\pi}\big[ \eta - \frac{\eta}{\sqrt{4(\omega^2-\nu^2)}} F(\frac12,\frac12,2;\eta)   \big]\ .
\eea
Since it is difficult to extract the general form of the relation $E=E(J_+,J_-)$, we consider the two extreme limit, i.e. short strings ($\eta\ll1$) and long strings ($1-\eta\ll1$). 

\subsubsection{Short strings}
We have to consider the limit $\eta\ll1$. In this case from (\ref{pps}) and (\ref{dfn}) we get
\bea
\omega^2-\nu^2&=&1+{\eta\over2}+O(\eta^2)\ ,\cr
k^2-\nu^2&=&\eta+O(\eta^2)\ .
\eea
From (\ref{pps}), the rescaled angular momenta ${\cal J}_+$ and ${\cal J}_-$ defined as in (\ref{defresc}) become
\bea
{\cal J}_+ &=& \nu\big[ 1-\frac12 \eta+O(\eta^2)\big]\ ,\cr
{\cal J}_- &=&\frac 12 \omega\big[ \eta +O(\eta^2)\big]\ .
\eea
Then the relation ${\cal E}({\cal J}_+,{\cal J}_-)$ assumes the following approximate form 
\bea
{\cal E}= {\cal J}_+ - {{\cal J}_-\over {\cal J}_+}\sqrt{1 + {\cal J}_+^2}\ ,
\eea
giving the BMN-type expansion ($J=J_+-J_-$)
\be
E\sim J - \frac{\lambda^2J_-}{2J^2}\ .
\ee

\subsubsection{Long strings}
We now consider the limit $1-\eta\ll1$. In this case from (\ref{pps}) and (\ref{dfn}) we have 
\bea\label{longs}
\omega^2-\nu^2&=&\frac{1}{\pi^2}\ln ^2(1-\eta)\big[ 1-8\ln2\ln^{-1}(1-\eta)+o(\ln^{-1}(1-\eta))\big] \ ,\cr
k^2-\omega^2&=&o(\ln^{-n}(1-\eta)) \quad \ {\rm for\ any}\ n>0\ ,
\eea
while the angular momenta become
\bea
{\cal J}_+ &=& -2\nu\ln^{-1}(1-\eta)\big[ 1+4\ln2\ln^{-1}(1-\eta)+o(\ln^{-1}(1-\eta))\big]\ ,\cr
{\cal J}_- &=& \omega\big[1+2\ln^{-1}(1-\eta)+o(\ln^{-1}(1-\eta)) \big]\ .
\eea
Then, a little bit of algebra results in\footnote{This formula reduces to the single charge solution of \cite{pons} in the ${\cal J_+}\rightarrow 0$ limit.}
\bea
{\cal E}=-{\cal J}_-\left[ 1-\frac{2}{\cal J_-}\sqrt{\frac1{\pi^2} +\frac14{\cal J_+}^2}+o(\ln^{-1}(1-\eta))  \right]\ ,
\eea
giving the expansion
\be
E\sim J + \frac{2\lambda^2}{\pi^2J_+}\ .
\ee

\subsection{Circular constant radii solutions}\label{sezcirc}

In this section we will consider another class of solutions derivable from the general approach presented in section \ref{sezinteg}. These  have constant radii and can be considered as the bMN analog of the circular multi-spin string solutions extensively studied in \cite{frolov2}. Again, we follow the general procedure outlined in \cite{frolov5,russo,tseyrew}.   
Let us now reconsider the effective action (\ref{inteff}) with non-zero integrals of motion $v_i$. The particular class of solutions we are interested in can be obtained by considering the case of constant $\Lambda$ and correspondingly of constant radii $r_i$.
From the periodicity condition (\ref{peco}) and the relation $\alpha_i^\prime=v_i/r_i^2$ one can easily relate the $v_i$'s to the $m_i$'s, i.e. $v_i=m_ia_i^2$. 
Then, we can always write the solution in the form
\bea
z^1+iz^2=a_1e^{im_1\sigma}\ ,\quad\quad z^3+iz^4=a_2e^{im_2\sigma}\ ,\quad\quad a_1^2+a_2^2=1\ ,
\eea 
and the following relations hold
\bea\label{rere}
&&w_i^2=m_i^2-\Lambda\ ,\qquad\qquad\qquad\quad\  a_1^2+a_2^2=1\ ,\cr
&& k^2=2(a_1^2w_1^2+a_2^2w_2^2)+\Lambda\ ,\quad\quad a_1^2w_1m_1+a_2^2w_2m_2=0\ .
\eea
The two nonzero spins are 
\bea\label{circspin}
J_-=\lambda {\cal J}_-=-\lambda a_1^2w_1\ ,\quad\quad J_+=\lambda {\cal J}_+=\lambda a_2^2w_2\ .
\eea
From the relations (\ref{rere}) we have that
\bea
{\cal E}^2=2\Big(-{\cal J}_{-}\sqrt{m_1^2-\Lambda}+ {\cal J}_{+}\sqrt{m_2^2-\Lambda}  \Big)+\Lambda\ .
\eea
The explicit expression of $E$ as a function of $(J_-,J_+,m_1,m_2)$  can be given in approximate form in the limit of very large total spin ${\cal J}={\cal J}_+ -{\cal J}_-$, as we have done in the folded string case. In fact, from (\ref{circspin}) and (\ref{rere}) it follows that 
\bea
\sum_{i=1,2}\frac{|{\cal J}_i|}{w_i}=1\qquad\Rightarrow \qquad\sum_{i=1,2}\frac{|{\cal J}_i|}{\sqrt{m_i^2-\Lambda}}=1\ .
\eea
In the large ${\cal J}$ limit we have that $\Lambda< 0$ with $|\Lambda|\gg 1$, so that we can make the following approximation
\bea
\sqrt{|\Lambda|}={\cal J}-\frac{1}{2{\cal J}}\Big(-m_1^2\frac{{\cal J}_-}{\cal J}+m_2^2\frac{{\cal J}_+}{\cal J}\Big)+\ldots\ .
\eea
Then, in the limit of large total rescaled spin ${\cal J}$, one obtains the following relation between the energy and the spins
\bea\label{energy-spincirc}
E=J\Big[1+\frac{\lambda^2}{2J^2}\Big(-m_1^2\frac{J_{-}}{J}+m_2^2\frac{J_+}{J}\Big)+\ldots\Big]\ .
\eea
As in the folded string case, this expression is regular in the effective coupling constant $\lambda^2/J^2$. 
In the following section we will perform a semiclassical analysis of this particular class of solutions. They allow to circumvent some subtleties in the quantization procedure which are present, for example, in the folded string case.
As a price to be paid for this, we will see that the solution is not stable for all the values of the parameters, some frequencies being imaginary.
But this is expected, the folded string being less energetic than the circular one with the same charges \cite{frolov4}.
We envisage that the quantization of the folded strings would give very similar results without instabilities.

\section{Semiclassical analysis of the circular solutions}\label{sezsemi}

In the coordinates $(\chi,\phi_-,\phi_-)$ our classical constant radii circular solutions has the following simple form
\bea\label{solcurv}
\sin\bar\chi=a_1\ ,\quad\quad \bar\phi_-=-(w_1\tau +m_1\sigma)\ ,\quad\quad \bar\phi_+=w_2\tau +m_2\sigma\ .
\eea
For $a_1,a_2\neq 0$ we can focus our attention on  the following thin neighborhood around $\chi=\bar\chi$ which contains the string world-sheet completely
\bea
&&\chi\in(\bar\chi-\epsilon,\bar\chi+\epsilon)\ ,\cr
&&\phi_-\in [0,2\pi]\quad\quad {\rm with}\ \phi_-\simeq \phi_-+2\pi\ ,\cr
&&\phi_+\in [0,2\pi]\quad\quad {\rm with}\ \phi_+\simeq \phi_++2\pi\ .
\eea   
This patch has the topology of an interval times a two-torus, with metric
\bea
ds^2=\frac{g_s}{m_0^2}[d\chi^2+\cos^2\chi d\phi_+^2+\sin^2\chi d\phi_-^2]\ .
\eea

To gain more insight in our solutions (\ref{solcurv}), we can make the following linear change of coordinates
\bea\label{change1}
v_1&=&\frac{1}{\sqrt{a_1^2 w_1^2+a_2^2 w_2^2}}\big( -a_1^2w_1\phi_-+a_2^2w_2\phi_+\big)\ ,\cr
v_2&=&\frac{1}{\sqrt{a_1^2 m_1^2+a_2^2 m_2^2}}\big( -a_1^2m_1\phi_-+a_2^2m_2\phi_+\big)\ .
\eea
In the new coordinates $(\chi,v_1,v_2)$, the solutions (\ref{solcurv}) take the form
\bea\label{solcurv2}
\chi(\tau,\sigma)&\equiv& \bar\chi \quad {\rm with}\ \sin\bar\chi=a_1\ ,\cr
v_1(\tau,\sigma)&\equiv& \bar v_1(\tau)= \tau\sqrt{a_1^2 w_1^2+a_2^2 w_2^2}\ ,\cr
v_2(\tau,\sigma)&\equiv& \bar v_2(\sigma)= \sigma\sqrt{a_1^2 m_1^2+a_2^2 m_2^2}\ .
\eea

\subsection{Pull-back of the bMN background on the classical solutions}
Let us then consider what is the form of the pull-back of the relevant fields on the solutions we are interested in.
The induced metric is conformally flat
\bea\label{indmetr}
P_{\rm string}[ds^2]=h_{\alpha\beta}d\xi^\alpha d\xi^\beta=\frac{g_s}{m_0^2}(a_1^2 m_1^2+a_2^2 m_2^2)
\big(-d\tau^2+d\sigma^2\big)\ .
\eea
On the world-sheet, we can make the following choice of the 10d vielbein (the underlined numbers refer to flat directions) 
\bea\label{spinviels}
&&e^{\ul0}=\frac{\sqrt{g_s}}{m_0\sqrt{a_1^2 m_1^2+a_2^2 m_2^2}}\big(k dt-\sqrt{a_1^2 w_1^2+a_2^2 w_2^2}dv_1\big)\ ,\cr
&&e^{\ul8}=\frac{\sqrt{g_s}}{m_0\sqrt{a_1^2 m_1^2+a_2^2 m_2^2}}\big(-\sqrt{a_1^2 w_1^2+a_2^2 w_2^2}dt+k dv_1\big)\ ,\cr
&&e^{\ul1,\ul2,\ul3}=\sqrt{\frac{g_s}{m_0^2}}dX^{1,2,3}\ ,\qquad\qquad e^{\ul4,\ul5,\ul6}=\sqrt{\frac{g_s}{m_0^2}}dY^{1,2,3}\ ,\cr
&&e^{\ul7}=\sqrt{\frac{g_s}{m_0^2}}d\chi\ ,\qquad\qquad \qquad \quad e^{\ul9}=\sqrt{\frac{g_s}{m_0^2}}dv_2\ .
\eea
Then, we have that the pull-back of this 10d vielbein on the world-sheet has only the following non-zero elements
\bea
P_{\rm string}[e^{\ul0}]=\sqrt{\frac{g_s}{m_0^2}}\sqrt{a_1^2 m_1^2+a_2^2 m_2^2}d\tau\ ,\qquad\quad 
P_{\rm string}[e^{\ul9}]=\sqrt{\frac{g_s}{m_0^2}}\sqrt{a_1^2 m_1^2+a_2^2 m_2^2}d\sigma\ .
\eea 
On the world-sheet, the RR field strength takes the form
\bea
F^{(3)}&=&-\frac{2}{m_0^2}\left(  \frac{m_0^2}{g_s}\right)^{\frac{3}{2}}\frac{a_2w_2}{a_1m_1}
          \Big[e^{\ul0}\wedge e^{\ul7}\wedge e^{\ul9}-\sqrt{1+\Big(\frac{a_1m_1}{a_2w_2}\Big)^2}e^{\ul7}
          \wedge e^{\ul8}\wedge e^{\ul9}\Big]+\cr
        && +\frac{b}{m_0^2}\left(  \frac{m_0^2}{g_s}\right)^{\frac{3}{2}}\frac{1}{\sqrt{a_1^2 m_1^2+a_2^2 m_2^2}}\Bigl[ (w_1+w_2)\frac{a_1^2m_1}{w_2}e^{\ul4}\wedge e^{\ul5}\wedge e^{\ul9}+\cr
                &&- (m_1+m_2)\frac{a_2^2w_2}{m_1}(e^{\ul0}\wedge e^{\ul4}\wedge e^{\ul5}+ \sqrt{1+\Big(\frac{a_1m_1}{a_2w_2}\Big)^2}e^{\ul4}\wedge e^{\ul5}\wedge e^{\ul8})+\cr
                && +a_1a_2\Bigl((w_1+w_2)(e^{\ul0}\wedge e^{\ul5}\wedge e^{\ul6}+ \sqrt{1+\Big(\frac{a_1m_1}{a_2w_2}\Big)^2}e^{\ul5}\wedge e^{\ul6}\wedge e^{\ul8})+\cr
            && + (m_1+m_2)e^{\ul5}\wedge e^{\ul6}\wedge e^{\ul9}\Bigr) \Bigr]  +\frac{b}{m_0^2}\left(  \frac{m_0^2}{g_s}\right)^{\frac{3}{2}} e^{\ul4}\wedge e^{\ul6}\wedge e^{\ul7}\ .
\eea
To end this section, when we will consider the fermionic sector we will need the explicit form of the non-zero components of the pull-back of the 10d spin connection 
\bea
\Omega_\tau^{\ul{07}}&=& -a_1a_2\frac{m_1^2-m_2^2}{\sqrt{a_1^2 m_1^2+a_2^2 m_2^2}}\ ,\cr
\Omega_\tau^{\ul{78}}&=& -a_1a_2\frac{m_1^2-m_2^2}{\sqrt{a_1^2 m_1^2+a_2^2 m_2^2}}\sqrt{1+\frac{a_1^2m_1^2}{a_2^2w_2^2}}\ ,\cr
\Omega_\tau^{\ul{79}}&=& -a_1a_2\frac{w_1m_1-w_2m_2}{\sqrt{a_1^2 m_1^2+a_2^2 m_2^2}}\ ,\cr
&&\cr
\Omega_\sigma^{\ul{07}}&=& -a_1a_2\frac{w_1m_1-w_2m_2}{\sqrt{a_1^2 m_1^2+a_2^2 m_2^2}}\ ,\cr
\Omega_\sigma^{\ul{78}}&=& -a_1a_2\frac{w_1m_1-w_2m_2}{\sqrt{a_1^2 m_1^2+a_2^2 m_2^2}}\sqrt{1+\frac{a_1^2m_1^2}{a_2^2w_2^2}}\ ,\cr
\Omega_\sigma^{\ul{79}}&=& -a_1a_2\frac{m_1^2-m_2^2}{\sqrt{a_1^2 m_1^2+a_2^2 m_2^2}}\ .
\eea
\subsection{Bosonic fluctuations}
In order to study the quadratic fluctuations around these solutions, we can use the Nambu-Goto formulation\footnote{In this case, thanks to the regularity of our solutions (the induced metric is conformally flat with constant conformal factor) the usual problems related to this formulation are not present.} as well as the Polyakov one. The calculations are very easy in the latter case, due to the simplifications related to the choice of the conformal gauge. 
We thus consider the following fluctuations of our fields  around the solutions (\ref{solcurv})
\bea
&&t=k\tau+{\delta t\over\sqrt\lambda}\ ,\quad\quad \phi_-=\bar\phi_-+{\delta\phi_-\over\sqrt\lambda}\ ,\quad\quad 
\phi_+=\bar\phi_++{\delta\phi_+\over\sqrt\lambda}\ ,\cr
&& \chi=\bar\chi+{\delta \chi\over\sqrt\lambda}\ ,\quad\quad X^a={\delta X^a\over\sqrt\lambda}\ ,\quad\quad Y^a={\delta Y^a\over\sqrt\lambda}\ .
\eea
In order to write the Lagrangian in a simple form, it is convenient to introduce new coordinates, inspired by the vielbein (\ref{spinviels}),
\bea
\d \tilde t &=& \frac{1}{\sqrt{a_1^2m_1^2+a_2^2m_2^2}}\Big( k \delta t -\delta v_1\sqrt{a_1^2w_1^2+a_2^2w_2^2}\Big)\ ,\cr
\d \tilde v_1 &=&  \frac{1}{\sqrt{a_1^2m_1^2+a_2^2m_2^2}}\Big(  -\delta t\sqrt{a_1^2w_1^2+a_2^2w_2^2}+k\delta v_1\Big)\ .
\eea

In the conformal gauge in which the world-sheet metric is fixed to be equal to the induced metric (\ref{indmetr}), the bosonic part of the Polyakov action is given by
\bea\label{Pac}
S_P&=&-\frac{\lambda}{4\pi}\int d\tau\int_0^{2\pi}d\sigma \big(-h_{\tau\tau}+h_{\sigma\sigma}\big)=\\
&=& -\frac{\lambda}{4\pi}\int d\tau\int_0^{2\pi}d\sigma\big[(-h_{\tau\tau}+h_{\sigma\sigma})_|+(-\delta h_{\tau\tau}+\delta h_{\sigma\sigma})_|+\frac12(-\delta^2 h_{\tau\tau}+\delta^2 h_{\sigma\sigma})_|+\ldots\big]\ .\nonumber
\eea
The conformal constraints $\ h_{\tau\tau}+ h_{\sigma\sigma}=0,\   h_{\tau\sigma}=0\ $ at leading order read
\bea\label{momenta}
&&-\pa_\tau\delta\tilde t +\pa_\sigma\delta v_2+\frac{2a_1a_2(m_1^2-m_2^2)}{\sqrt{a_1^2m_1^2+a_2^2m_2^2}}\delta\chi=0\ ,\cr
&&\pa_\tau\delta v_2 -\pa_\sigma\delta\tilde t +\frac{2a_1a_2(w_1m_1-w_2m_2)}{\sqrt{a_1^2m_1^2+a_2^2m_2^2}}\delta\chi=0\ .
\eea
After imposing these constraints to the Polyakov action we get the same result we would have obtained in the Nambu-Goto formulation. The fields $\d \tilde t$ and $\d v_2$ decouple completely from the action and represent unphysical degrees of freedom, and the bosonic action for the remaining eight transverse fluctuation is found, after some algebra, to be equal to
\bea\label{redac}
S^{fl}&=& -\frac{1}{4\pi}\int d\tau\int_0^{2\pi}d\sigma \,\Big\{\eta^{\alpha\beta}\pa_\alpha \d X^a\pa_\beta
\d X^a+ \eta^{\alpha\beta}\pa_\alpha \d Y^a\pa_\beta
\d Y^a+\eta^{\alpha\beta}\pa_\alpha \d \chi\pa_\beta\d \chi+\cr 
&&\qquad +\eta^{\alpha\beta}\pa_\alpha \d \tilde v_1\pa_\beta\d \tilde v_1+
A(\d \chi)^2
+B\d \chi
[(m_2^2-m_1^2)\pa_\tau\d \tilde v_1 + (w_1 m_1-w_2m_2)\pa_\sigma\d \tilde v_1]+\cr
&&\qquad +M \d Y^a\d Y^a +((\d Y^1)^2+(\d Y^2)^2)M_1 + 
(\d Y^1\partial_\t \d Y^2 -\d Y^2\partial_\t \d Y^1)M_2 +\cr
&&\qquad +(\d Y^1\partial_\s \d Y^2 -\d Y^2\partial_\s \d Y^1)M_3 +(\d Y^2\partial_\t \d Y^3 -\d Y^3\partial_\t \d Y^2)M_4 +\cr
&&\qquad +
(\d Y^2\partial_\s \d Y^3 -\d Y^3\partial_\s \d Y^2)M_5\Big\}\ ,
\eea
where
\bea
A&=&-4\frac{a_1^2a_2^2}{a_1^2m_1^2+a_2^2m_2^2}[(m_1^2-m_2^2)^2-(w_1m_1-w_2m_2)^2]\ ,\cr
B&=&\frac{4k a_1a_2}{\sqrt{a_1^2m_1^2+a_2^2m_2^2}\sqrt{a_1^2w_1^2+a_2^2w_2^2}}\ ,\cr
M&=&2\Big(\frac{b^2}{4} +\frac13\Big)(a_1^2m_1^2+a_2^2m_2^2)\ ,\cr
M_1&=&(1-\frac b2)[(m_1-m_2)^2-(w_1-w_2)^2]\ ,\cr
M_2&=&(2-\frac b2)(w_2-w_1)-\frac b2 (a_2^2-a_1^2)(w_1+w_2)\ ,\cr
M_3&=&(-2+\frac b2)(m_2-m_1)+\frac b2 (a_2^2-a_1^2)(m_1+m_2)\ ,\cr
M_4&=& ba_1a_2(w_2+w_1)\ ,\cr
M_5&=& -ba_1a_2(m_2+m_1)\ .
\eea

\subsection{The fermionic sector}
\vspace{6pt}
The term of the superstring action quadratic in the fermions is given by (see also Appendix A) 
\bea\label{fermac}
S_{\rm (ferm)}=\frac{i}{4\pi\alpha^\prime}\int d\tau\int_0^{2\pi}d\sigma{ [\sqrt{-h}h^{\alpha\beta}\delta^{IJ}-\epsilon^{\alpha\beta}(\sigma_3)^{IJ}]
\bar\theta^I\rho_\alpha(D_\beta \theta)^J}\ ,
\eea
where $h_{\alpha\beta}$ is the induced metric, $\rho_\alpha=\partial_\alpha x^m e_m^a \Gamma_a$ and $D_\alpha$ is the pullback on the world-sheet of
\bea
D_m &=&\nabla_m+\frac1{8\cdot 3!} e^{\Phi}{F}^{(3)}_{abc}\Gamma^{abc}\Gamma_m\sigma_1\ ,\cr
\nabla_m&=&\partial_m +\frac14 \Omega_{mab}\Gamma^{ab}\ .
\eea
After imposing the $\kappa$-symmetry gauge
\bea
\theta^1=\theta^2=\theta\ ,
\eea
and performing the usual rescalings in order to factorize the dependence on $\lambda$
\bea
ds^2\rightarrow \frac{g_s}{m_0^2}ds^2\quad,\quad e^a\rightarrow \left(\frac{g_s}{m_0^2}\right)^{\frac12}e^a
\quad,\quad \theta\rightarrow \left(\frac{g_s}{m_0^2}\right)^{\frac14}{\theta\over\sqrt\lambda}\ ,
\eea
a straightforward calculation shows that the fermionic string action (\ref{fermac}) takes the form
\bea\label{feracfin}
S_{\rm (ferm)}&=&
\frac{i}{2\pi}\int  d\tau\int_0^{2\pi}d\sigma \bar\theta\Big\{\sqrt{a_1^2 m_1^2+a_2^2 m_2^2}\big(-\Gamma_{\ul 0} \partial_\tau+
\Gamma_{\ul 9} \partial_\sigma\big)+\cr
&&+\frac{a_1a_2}{2}\Big[\sqrt{1+\frac{a_1^2m_1^2}{a_2^2w_2^2}}\Big((m_1^2-m_2^2)\Gamma_{\ul{078}}-(w_1m_1-w_2m_2)\Gamma_{\ul{789}}\Big)+(w_1m_2-w_2m_1)\Gamma_{\ul{079}}\Big]\cr
&&+ \frac{b}{4}\sqrt{a_1^2 m_1^2+a_2^2 m_2^2}\Big[\sqrt{1+\Big(\frac{a_1m_1}{a_2w_2}\Big)^2}\Big((m_1+m_2)\frac{a_2^2w_2}{m_1}\Gamma_{\ul{458}} -a_1a_2(w_1+w_2)\Gamma_{\ul{568}} \Big)+\cr
&&- \sqrt{a_1^2 m_1^2+a_2^2 m_2^2}\Gamma_{\ul{467}} \Big]
\Big\}\theta\ .
\eea
\subsection{One loop correction to the energy}
In order to calculate the one-loop correction to the energy/charge relation (\ref{energy-spincirc}) in the large $k$ limit, we will approximate
the series over $n$ of the frequencies with integrals\footnote{Let us note that in literature one often finds another way of calculating the zero-point energy. It is based on {\it renormalizing} \`a la Casimir the bosonic and fermionic series separately, and eventually sum the two contributions. The renormalization amounts in subtracting to the series exactly their integral approximation (the one we are going to calculate). Obviously, since the latter is the dominant contribution at large $k$, using this procedure one finds that in this limit the corrections are always exponentially vanishing, by construction. This would be true for {\it every} semiclassical solution studied in this context. As such, the discussion about the subleading behavior, at large $k$, of the quantum corrections would be trivial, the quantum contributions being subleading by construction. But, since there is no convincing argument in favor of this renormalization procedure (see the discussions in \cite{noistab} and appendix D of \cite{univers}), we will not use it and we will simply calculate the (finite) vacuum energy approximating it with integrals.} in $x\equiv n/k$ \cite{frolov1,noi0,noistab,frolov3}.
We will then derive the characteristic frequencies for the string fields in the large $k$, fixed $x$ limit.
The form of the metric (\ref{metrnew2}), where the charges $J_+,\, J_-\,$ correspond to Killing directions, makes it straightforward to show that, since for our solutions $t=k\tau$, the world-sheet vacuum energy we are going to calculate corresponds to $k$ times the leading quantum correction to the space-time energy/charge relation (see appendix A of \cite{frolov1}).  
In appendix \ref{appendixb} we show that the theory is UV finite and briefly discuss its stability.

The action (\ref{redac}) accounts for the eight physical bosonic fields.
Three of them, the $\delta X^a$ ones, are massless as expected.
Their frequencies are therefore $\omega_b^i=n=kx,\,\,i=1,2,3$.

Diagonalizing the equations of motion of the $\delta \tilde{v}_1,\, \delta \chi$ fields 
we find the equation for the frequencies
\be\label{freqs1}
(2\omega_b^2-2n^2-A)^2-A^2-B^2[(m_2^2-m_1^2)\omega_b+(w_1m_1-w_2m_2)n]^2=0\ .
\ee
In the large $k$, fixed $x$ limit the two frequencies have then the form 
\be
\omega_b \sim k(\sqrt{x^2+1}\pm 1) \pm  \frac{|x|(m_1+m_2)}{\sqrt{x^2+1}} + O(\frac1k)\ .
\ee
For all the bosonic and fermionic frequencies we do not write down the (involved) expressions of the $k^{-1}$ coefficients
because we will not need them in the following.

Diagonalizing the equations of motion of the $\delta Y^a$ fields we get the following equation
\bea\label{freqy}
(\omega_b^2-n^2-M-M_1)^2=(n M_5+\omega_b M_4)^2+(n M_3+\omega_b M_2)^2-\frac{M_1(n M_5+\omega_b M_4)^2}{\omega_b^2-n^2-M}\ .
\eea
The large $k$, fixed $x$ limits of the three frequencies read 
\bea
\omega_b &\sim& k(\sqrt{x^2+\frac{b^2}{4}}\pm \frac b2) \pm \frac{|x|(b/4-1)(m_1+m_2)}{\sqrt{x^2+\frac{b^2}{4}}} + O(\frac{1}{k})\ ,\cr
\omega_b &\sim& k|x| + O(\frac{1}{k})\ .
\eea
Thus, the sum of the bosonic frequencies in the limit gives\footnote{Note that the $O(k^0)$ term vanishes.}
\be\label{sumbosk}
\sum \omega_b = k\Big( 4|x| +2\sqrt{x^2+\frac{b^2}{4}}+2\sqrt{x^2+1} \Big) + O(\frac{1}{k})\ .
\ee
Remarkably, this is exactly the expression one finds in the Penrose limit calculation for the bMN background \cite{abc}.
The frequencies are slightly different but retain the main features: we have three massless world-sheet bosons describing
the flat three special directions, two modes\footnote{Remember that we rescaled all our quantities factorizing
out $m_0$, which is the unit measure for the masses. Reestablishing it simply amounts in the rescaling $k \rightarrow k/m_0$.} (the ``universal sector'' $v$ bosons in the notation of \cite{pandostras,abc}) which describe the three-sphere excitations, and two $b$-dependent modes (the $u$ bosons \cite{pandostras,abc}) which, together with a
 fourth massless mode (the $z$ field \cite{pandostras,abc} whose masslessness is peculiar of the (b)MN background), describe
 the fibrating $\mathbb{R}^3$ geometry.
This pattern of modes is universal in the Penrose limit of confining backgrounds \cite{univers}.
\\

The equation of motion for the fermions coming from the action (\ref{feracfin}) is
\be\label{fereqmot}
[(-\Gamma_{\ul0}\partial_\tau + \Gamma_{\ul9}\partial_\sigma) + C \Gamma_{\ul{078}} + D\Gamma_{\ul{789}} + E\Gamma_{\ul{079}}+ F\Gamma_{\ul{458}}+ G\Gamma_{\ul{568}}+ H\Gamma_{\ul{467}}]\theta=0\ ,
\ee
where 
\bea
C&=&\frac{a_1a_2}{2\sqrt{a_1^2 m_1^2+a_2^2 m_2^2}}\sqrt{1+\frac{a_1^2m_1^2}{a_2^2w_2^2}}(m_1^2-m_2^2)\ ,\cr
D&=&-\frac{a_1a_2}{2\sqrt{a_1^2 m_1^2+a_2^2 m_2^2}}\sqrt{1+\frac{a_1^2m_1^2}{a_2^2w_2^2}}(w_1m_1-w_2m_2)\ ,\cr
E&=&\frac{a_1a_2}{2\sqrt{a_1^2 m_1^2+a_2^2 m_2^2}}(w_1m_2-w_2m_1)\ ,\cr
F&=&\frac{b}{4}\sqrt{1+\Big(\frac{a_1m_1}{a_2w_2}\Big)^2}(m_1+m_2)\frac{a_2^2w_2}{m_1}\ ,\cr
G&=&-\frac{b}{4}\sqrt{1+\Big(\frac{a_1m_1}{a_2w_2}\Big)^2}a_1a_2(w_1+w_2)\ ,\cr
H&=&-\frac{b}{4}\sqrt{a_1^2 m_1^2+a_2^2 m_2^2}\ .
\eea
Applying to the right of (\ref{fereqmot}) the operator
\be
(\Gamma_{\ul9}\partial_\tau - \Gamma_{\ul0}\partial_\sigma) + C\Gamma_{\ul{978}} + D\Gamma_{\ul{078}} + E\Gamma_{\ul{7}}+ F\Gamma_{\ul{45809}}+ G\Gamma_{\ul{56809}}+ H\Gamma_{\ul{46709}}\ ,
\ee
we get the equation
\bea
&&[\ -\partial^2_\tau + \partial^2_\sigma + 2(C\partial_{\sigma}+ D\partial_{\tau})\Gamma_{\ul{0978}}  - C^2 + D^2  - E^2 - F^2 - G^2 - H^2 +\cr
&&\qquad \qquad\qquad \  -2EF\Gamma_{\ul{097845}}-2CG\Gamma_{\ul{097856}}
+2FH\Gamma_{\ul{5678}}-2GH\Gamma_{\ul{4578}}\ ]\ \theta=0\ .
\eea
Choosing the four dimensional Weyl representation for the gamma matrices $\Gamma_{\ul{0,9,7,8}}$, for which $\Gamma_{\ul{0978}}={\rm diag}(-i,-i,i,i)$ and $\Gamma_{\ul{78}}={\rm diag}(-i,i,-i,i)$, and the two dimensional $\Gamma_{\ul{56}}={\rm diag}(-i,i)$, $\Gamma_{\ul{45}}={\rm antidiag}(-i,-i)$, the equation for the eight frequencies finally read
\bea\label{freqfer}
(\omega_f^2-n^2- C^2 + D^2 - E^2 - F^2 - G^2 - H^2 + 2Cn + 2D\omega_f)^2\cr 
-4(EF+GH)^2 -4(-EG+FH)^2=0\ ,\cr
(\omega_f^2-n^2- C^2 + D^2 - E^2 - F^2 - G^2 - H^2 + 2Cn + 2D\omega_f)^2\cr 
-4(EF-GH)^2 -4(EG+FH)^2=0\ ,\cr
(\omega_f^2-n^2- C^2 + D^2 - E^2 - F^2 - G^2 - H^2 -2Cn -2D\omega_f)^2\cr 
-4(EF-GH)^2 -4(EG+FH)^2=0\ ,\cr
(\omega_f^2-n^2- C^2 + D^2 - E^2 - F^2 - G^2 - H^2 - 2Cn - 2D\omega_f)^2\cr 
-4(EF+GH)^2 -4(EG-FH)^2=0\ .
\eea
The large $k$, fixed $x$ limit for the first four frequencies is
\be
\omega_f \sim k \Big[\frac12+\sqrt{\Big(\frac12 \pm \frac{b}{4}\Big)^2+x^2} \Big] - \frac{|x|(m_1+m_2)}{2\sqrt{(\frac12 \pm \frac{b}{4})^2+x^2}} + O(\frac{1}{k})\ ,
\ee
while for the last four it reads
\be
\omega_f \sim k \Big[-\frac12+\sqrt{\Big(\frac12 \pm \frac{b}{4}\Big)^2+x^2} \Big] + \frac{|x|(m_1+m_2)}{\sqrt{(\frac12 \pm \frac{b}{4})^2+x^2}} + O(\frac{1}{k})\ .
\ee
The sum of the fermionic frequencies is then\footnote{Again, the $O(k^0)$ term vanishes.}
\be
\sum \omega_f = k\Big[ 4\sqrt{\Big(\frac12 + \frac{b}{4}\Big)^2+x^2} + 4\sqrt{\Big(\frac12 - \frac{b}{4}\Big)^2+x^2} \Big] + O(\frac{1}{k})\ ,
\ee
which is {\it not} equal to the bosonic one (\ref{sumbosk}) already at level $k$.
Again, even if the single frequencies are different from the pp-wave ones \cite{abc}, their sum is exactly the same. 

As a result, the leading one-loop sigma-model contribution to the string zero-point energy (and so to the energy/charge relation)
in the large $k$ limit is non vanishing
\bea
E_1&=&{1\over 2k}\sum_{n=-\infty}^{\infty}\left[\sum_b \omega_b(n) - \sum_f \omega_f(n)\right] \cr
&\approx& -{k\over2}\Big[\frac{b^2}{4}\log{\frac{b^2}{4}}-2\Big(\frac12 + \frac{b}{4}\Big)^2\log{\Big(\frac12 + \frac{b}{4}\Big)^2}
-2\Big(\frac12 - \frac{b}{4}\Big)^2\log{\Big(\frac12 - \frac{b}{4}\Big)^2}\Big]\ ,
\label{E_1}
\eea
where in the last step we have approximated the series with integrals.
It is of course the same zero-point energy of the Penrose limit of the bMN theory in the large charge limit \cite{abc}.
Calling $m_b$ and $m_f$ the masses as they appear in the sum of the frequencies (for example, we have for the bosons
$\sqrt{x^2+m_b^2}$ with $m_b=(0,1,b/2)$), the zero point energy (\ref{E_1}) has the ``universal'' form
\be
E=-{k\over4}[\sum_b m_b^2\log{m_b^2}- \sum_f m_f^2\log{m_f^2}]\ .
\ee
It is exactly the same form one finds in the large class of plane-wave backgrounds studied in \cite{noistab}.
Moreover, the finiteness of $E_1$ is guaranteed by the relation
\be
\sum_b m_b^2=\sum_f m_f^2\ .
\ee
In the Penrose limit case \cite{pandostras,abc,sonne,univers} and on $AdS_5 \times S^5$ \cite{frolov3} it is a consequence
of the supergravity field equations.
In our case, it can be viewed as the indication that the quantization procedure we adopted is consistent.

Let us comment on the relation (\ref{E_1}) in some detail. First we notice that
the one-loop contribution to the zero-point energy is linear in $k$ and it is always {\it negative} for $b\in(0,2/3]$ \cite{noistab,abc}.
The fact that the circular string gives the same one-loop correction as the point-like string of the Penrose limit can be
understood as follows.
In section \ref{sezsemi} we argued that the circular string is wrapped on a two-torus, with every point of the string moving orthogonally
to the string displacement.
Now, the boost of a single point is exactly the Penrose limit theory, having the zero-point energy (\ref{E_1}).
Since, as can be seen from (\ref{cariche}), for our case of constant radii the angular momentum density is constant along the string, the total zero-point energy 
is nothing than the average of the same $E_1$ for every point of the string, i.e. $E_1$ itself.

The very same reasoning can be applied to the circular string configurations on the five-sphere of $AdS_5 \times S^5$.
In that case too one obtains an average of the Penrose limit theory for every point of the string to leading order in the large $k$ expansion.
The difference is that
on this plane-wave 
the zero-point energy is vanishing, due to the underlying supersymmetry, 
and so the order $k$ one-loop correction 
is zero for the circular strings.

Curiously, although outside the range of the values for which the correspondence with the field theory is sensible,
the value $b=2$ gives a zero for $E_1$.
This value of $b$ corresponds to a regular geometry which interpolates between a $b$-deformed MN background for small radius
and a factorized geometry $\mathbb{R}^4\times\mathbb{R}^3\times S^3$, with unit sphere, for large radius \cite{gubser}
(in the notation of that paper this critical value is $b=1/2$).
Again, in the large $k$ limit, the zero-point energy at order $k$ is zero because the zero-point energy on the pp-wave is vanishing in this theory.
The latter property is due to the fact that on these pp-wave backgrounds \cite{pandostras,abc} the world-sheet theory becomes supersymmetric precisely for $b=2$, as can be seen from the spectrum \cite{abc}.

\section{Supersymmetry analysis}\label{sezsusy}

The four supersymmetries preserved by the MN background are broken by our string solitons. This can be understood in the following way. Suppose that $\epsilon$ is the killing spinor of the MN solution. It satisfies the Killing spinor equations
\bea
D_m\epsilon=0\ ,\quad\quad \Delta\epsilon=0\ ,
\eea 
where $D_m$ and $\Delta$ are the operators that enter the supersymmetry transformation of the gravitino and the dilatino, whose explicit form is given in appendix \ref{appA}. We are using double spinor notation, i.e. $\epsilon=(\epsilon^1,\epsilon^2)$. Then, using the decompositions (\ref{operatorsIIB}) and (\ref{operatorsIIB2}) and posing
\bea
&&D_m^{(0)}=\hat D_m^{(0)}\otimes {\bf 1}_{2\times 2}\ ,\quad \quad W_m=\hat W_m \otimes \sigma_1\ ,\cr 
&&\Delta^{(1)}=\hat \Delta^{(1)}\otimes {\bf 1}_{2\times 2}\ ,\quad \quad \Delta^{(2)}=\hat \Delta^{(2)} \otimes \sigma_1\ ,
\eea
we can make explicit the killing equation in the following form
\bea\label{respi}
&&\hat D_m^{(0)}\epsilon^1+\hat W_m\epsilon^2=0\ ,\quad\quad \hat D_m^{(0)}\epsilon^2+\hat W_m\epsilon^1=0\ ,\cr
&& \hat \Delta^{(1)}\epsilon^1+\hat \Delta^{(2)}\epsilon^2=0\ ,\quad\quad  \hat \Delta^{(1)}\epsilon^2+\hat \Delta^{(2)}\epsilon^1=0\ .
\eea
Due to the Poincare isometry group of the four flat directions, these conditions must have a unique solution up to a Lorentz transformation that permits to obtain the four independent killing spinors. From (\ref{respi}) it is clear  that if $\epsilon=(\epsilon^1,\epsilon^2)$ is a killing spinor, then  $\tilde\epsilon=(\epsilon^2,\epsilon^1)$ is a killing spinor too. But the second cannot be in general obtained from the first one by a Lorentz transformation unless $\epsilon^1$ and $\epsilon^2$ are proportional\footnote{This can be understood by choosing the four dimensional Weyl representation for the killing spinors. Then, in two-components formalism, the general 4d Lorentz transformation is realized by the action of an unimodular $2\times 2$ complex matrix $M\in SL(2,\mathbb{C})$. Taking $\epsilon^1=(1,0)$ and imposing $M^2\epsilon^1=\epsilon^1=1$ with $M\in SL(2,\mathbb{C})$, it is simple to see that $\epsilon^2=M\epsilon^1=\pm \epsilon^1$.}. We then conclude that
\bea\label{killcond}
\epsilon^1\propto \epsilon^2\ .
\eea
This is indeed consistent with the hypothesis $\epsilon^1=\epsilon^2$ made in \cite{nunez} in the calculation of the explicit form of the supersymmetries of the MN solution. Let us now consider the GS superstring action on the MN background, whose general form is given in appendix \ref{appA}. The chiral operator appearing in the kappa-symmetry transformation of the GS superstring is of the form (see appendix \ref{appA})
\bea
\Gamma_{F1}=\frac{\epsilon^{\alpha\beta}}{2\sqrt{-h}}\Gamma_{\alpha\beta}\sigma_3=\hat\Gamma_{F1}\otimes\sigma_3\ .
\eea
Then, a (nondegenerate) string configuration preserve the MN supersymmetry $\epsilon$ if $(1+\Gamma_{F1})\epsilon=0$, or more explicitly
\bea\label{kappacond}
(1+\hat\Gamma_{F1})\epsilon^1=0\ ,\quad \quad (1-\hat\Gamma_{F1})\epsilon^2=0\ .
\eea
Since $(1\pm\hat\Gamma_{F1})$ are orthogonal projections, the conditions (\ref{killcond}) and (\ref{kappacond}) cannot be contemporaneously satisfied. 
In the point-like degenerate case supersymmetry breaking is still present \cite{pandostras}: even if the pp-wave background preserves sixteen supersymmetries, these do not give a supersymmetric spectrum, since they are not linearly realized. As we have seen in the previous sections, in the large $k$ limit the circular solution behaves like a bunch of point-like, non-interacting strings. As a consequence, supersymmetry is not restored in this limit. 
In the case of $b=2$ the supersymmetry {\it is} linearly realized on the pp-wave, and 
it would be very interesting to explicitly verify if this is the case also
in the $J\rightarrow \infty$ limit of our circular strings\footnote{We thank T. Mateos for a discussion on this issue.}.

\section{Multicharged hadrons in the dual picture}\label{sezmulti}

Let us now discuss the hadrons dual to the string solutions we have found.
We have analyzed configurations of spinning strings on the internal three-sphere, carrying two charges.
These are generalizations of the point-like string, carrying a single charge, studied in the Penrose limit in \cite{pandostras}
for the MN case and in \cite{abc} for the bMN one.
In the latter paper, an interpretation in the dual field theory for some string modes were given.
We will now extend that discussion to the two-charge case.

The dual field theory is a four dimensional (softly broken for the bMN case) (S)YM coupled to a (infinite) tower of KK fields \cite{mn}.
It comes from the world-volume low energy theory of five-branes upon wrapping the latter on a two-sphere inside a
(non compact) Calabi-Yau manifold.
The latter condition translates in the field theory in a twisting in the two-sphere directions, an operation
which allows to retain \none supersymmetry (eventually broken).

Among all the fields in the theory, the lightest KK scalars are the interesting ones for our purpose\footnote{For a discussion
of similar states in general confining backgrounds, see \cite{univers}.}, since the gauge degrees of freedom are uncharged under
the internal symmetries.
They come from the four scalars of the five-brane theory parameterizing the four directions transverse to the branes in flat space.
Upon wrapping of the branes, these scalars get all the same mass, proportional to $m_0$.
Moreover, being charged under the R-symmetry $SO(4)$ before wrapping, after wrapping and twisting they are charged under
the two surviving $U(1)_{l,r}$ inside $SU(2)_{l,r}$ from $SO(4)\sim SU(2)_{l}\times SU(2)_{r}$.
The combinations of these two U(1)'s are exactly the ones corresponding to our two charges.

It is not difficult to see that the scalars are in the bifundamental of the two groups, so that we can name them according to
these two charges as $A_{\pm,\pm}$, the first (second) entry referring\footnote{We use the same notation of \cite{abc}, in which
the second entry refers to the charge of the anti-fundamental representation, so ``+'' means charge $-1/2$ and ``$-$'' charge $+1/2$.} to $U(1)_l$ ($U(1)_r$) \cite{abc}.
In the single charge $J_+=J_l+J_r$ case studied in \cite{pandostras,abc}, the state of the form ${\Tr} (A_{+-})^{p}|0\rangle_{FT}$, $p\gg 1$, is identified with the string ground state.
In fact, since each component has charge $J_+=1$ and mass\footnote{This is not calculable in perturbative field theory. Rather, it is a prediction
of the string theory on the plane wave.} $m_0$, the energy/charge relation reads $E=m_0J_+$.
This corresponds to the energy of the string ground state $H=E-m_0J_+=0$, up to the zero-point energy correction.
The origin of the latter is unclear in field theory.
For some considerations, we refer to \cite{noi0,abc,univers,noitalk} and to the end of the present discussion.
Let us stress that the duality we are talking about is {\it not} between string states and operators, rather it is a correspondence
between string states and {\it hadrons}, named ``annulons'' in \cite{pandostras}.

In the two-charge case we are considering in this paper, together with the charge $J_+$, under which the scalar $A_{-+}$ has charge
$-1$ and the other two scalars $A_{++},\,A_{--}$ are uncharged, we have the charge $J_-=J_r-J_l$, under which $A_{+-},\,A_{-+}$ are uncharged, while $A_{++},\,A_{--}$ have charges $-1,\,+1$ respectively.
Thus, we immediately conclude that the string ground states are dual to hadrons which we will symbolically write as ${\Tr} [(A_{+-})^{p}(A_{++})^{q}]|0\rangle_{FT}$ ($p,q\gg 1$).
The total classical mass of the latter is $E=m_0(p+q)$, while they have $J_+,\,J_-$ charges $p,\,-q$.
As a consequence, they satisfy the relation $E=m_0(J_+-J_-)\equiv m_0J$.

This is the value of the hadron energies in the limit of infinite $\lambda,\ J$, with $\lambda/ J\rightarrow 0$. Specific string configurations with $J_+,J_-$ angular momenta, will provide specific informations on different ``internal'' structures of 
the above hadrons. Let us focus on the semiclassical circular solutions examined in the previous 
sections. The energy of the string ground state up to one-loop order, in the limit of large $J/\lambda$ is given by
\be\label{etot}
E=m_0J\Big[1+\frac{\lambda^2}{2J^2}\Big(-m_1^2\frac{J_{-}}{J}+m_2^2\frac{J_+}{J}\Big)+\ldots\Big]-m_0J\Big[{Z^2(b)\over\lambda}+\ldots\Big]\ ,
\ee 
where
\be
2Z^2(b)\equiv\frac{b^2}{4}\log{\frac{b^2}{4}}-2\Big(\frac12 + \frac{b}{4}\Big)^2\log{\Big(\frac12 + \frac{b}{4}\Big)^2}
-2\Big(\frac12 - \frac{b}{4}\Big)^2\log{\Big(\frac12 - \frac{b}{4}\Big)^2}\ .
\ee
The first terms in (\ref{etot}) originate from the large $J$ expansion of the classical energy (\ref{energy-spincirc}), while the last ones account for the one-loop zero-point energy evaluated in the large $k$ limit, i.e. formula (\ref{E_1}). We see that these terms have different nature. 

The terms coming from the classical energy are corrections in positive integer powers of the effective coupling $\lambda^2/J^2$. Moreover they depend on the string winding numbers, which possibly account for some internal (spin-chain like) structure of the dual hadrons and which are remnant of  
the particular string solutions we are analyzing. 
The limit of large $\lambda, J$ and small $\lambda^2/J^2$ we are taking 
can be read as an ``effective perturbative'' regime. Let us recall that 
$\lambda=Ne^{\Phi_0}$ is equal to the ratio  between the string 
tension and the squared mass of the hadron constituents $T_s/m_0^2$. So, 
we can speculate that taking the effective perturbative regime $T_s\ll Jm_0^2$ means considering a limit such that the contribution to the energy coming from the binding (which $T_s$ accounts for) of the hadron constituents is much smaller than the one coming from the masses of the constituents. 

The leading order corrections coming from the one-loop zero-point energy scale as $1/\lambda$ and as such they could be read as non perturbative (in $\lambda$) corrections to the mass of the single constituents of the hadrons.

Finally let us remember \cite{pandostras} that the low energy motion of our multicharged heavy states can  be described by an effective nonrelativistic particle moving in the three special dimensions (corresponding to the massless modes $\delta X^a$).

\section{Conclusions}\label{sezconc}
Let us summarize the key results we have found in this paper. We analyzed folded and circular string solutions on the asymptotic of the (b)MN background (but our conclusions about the classical configurations apply to the (b)KS case as well) corresponding to the low energy regime of a confining ${\cal N}=1$ (eventually softly broken) supersymmetric gauge theory coupled to an infinite tower of massive KK adjoint fields. 
String theory can be used to predict the energy/charge relations for hadrons formed by bound states of the latter. These relations can be easily studied in the limit of large charges, where a semiclassical analysis is possible. 
The classical string energy in the limit results to be proportional to the total charge, up to ``regular'' (i.e. positive powers in the effective coupling $\lambda^2/J^2$) corrections. 
Their presence is hard to reproduce in the dual field theory, due to the fact that what we are really examining is the strong coupling regime of a confining gauge theory. Nevertheless a possible interpretation of these corrections, as due to the collective binding energy of the hadron, can be given.

The semiclassical analysis of the above solutions give more informations on the energy/charge relations. 
We studied the case of circular solutions, where we were able to obtain the small effective coupling expansion for the one-loop sigma model contribution $E_1$ to the energy. As a first result we found that the leading term in $E_1$ is the same that would have been obtained by considering the zero-point energy of the string on a generalized Penrose limit of the (b)MN background\footnote{In the analogous case of multispinning circular solutions on the $S^5$ of the $AdS_5\times S^5$ background the same leading term is in fact zero (as the BMN zero-point energy).}. This term gives a correction to the classical energy/charge relation which scales as $1/\lambda$ and so is subleading with respect to the classical one (since we are assuming $\lambda\gg 1$). Nevertheless it is not suppressed in the limit of infinite total charge.
This  happens in general provided that the world-sheet supersymmetries are not linearly realized in the limit. 
This leading order quantum correction could be possibly interpreted as a non perturbative self renormalization of the mass of each single constituent of the dual hadron.

The main difference between the multispinning circular or folded string solutions and the point-like ones (which are semiclassically described by strings on pp-waves) is that the latter receive only the above kind of non perturbative, non analytic corrections to the energy/charge relation. At the classical level, in fact, string theory simply gives that $E$ is proportional to $J$ with no corrections to this relation. Thus they give different informations on the dual multiparticle bound states. 

Let us stress that our results will generalize to all the regular backgrounds dual to confining gauge theories present in literature\footnote{See \cite{Alishahiha:2004vi} for another example.}. In fact in the IR regime their form is of the universal form ${\mathbb R}^{d,1}\times{\mathbb R}^{q}\times S^{9-d-q} $. 
The classical configurations are then the $S^{9-d-q}$-versions of the ones studied in \cite{frolov5,russo,tseyrew} for $d+q=4$.
The quantization of the quadratic fluctuations can be performed along the lines of our section \ref{sezsemi}.  

It would be very interesting to study the folded string solutions at the semiclassical level. These are likely to give stable configurations and it would be nice to understand the one-loop corrections, and in particular to study the sign of the zero-point energy.

Finally, let us compare our results with the $AdS_5\times S^5$ ones.
In the latter case, whenever the classical solution has a spin on $S^5$, the expansion of the classical energy/charge relation is regular in $\lambda_t/J^2$. 
Since a lot of these configurations have been successfully compared with field theory computations, the common believe is that this regularity implies the vanishing of the quantum corrections in the asymptotic $J \rightarrow \infty$ limit.
In \cite{tmm1} it was claimed that this is due to the recovered supersymmetry (this was demonstrated for the circular strings on $S^5$) in the limit.
Nevertheless, there are examples of asymptotically non-BPS solutions whose classical results agree with the field theory computations \cite{mina}.
So, in \cite{tmm2} it was suggested that the possibility of successfully perform such comparisons should depend on the fact that the strings become effectively tensionless in the $J \rightarrow \infty$ limit.
The latter property is true because their size is bounded above by the radius of the five-sphere, so that the main contribution to the energy is the kinetic one.

The configurations we studied are non supersymmetric, even in the $J \rightarrow \infty$ limit.
Of course, the classical bosonic solutions do not mind about supersymmetry.
Since we have strings sitting at $\rho=0$, they see, at the classical level, only the geometry ${\mathbb R}^{3,1}\times S^3$.
The corresponding classical energy/charge relation is regular.
This could be interpreted with the criterion above about the effective vanishing tension of these strings, since at this level the three-sphere radius is fixed.
Note that the same criterion works for the other spinning strings on the MN background studied in \cite{pons}: the configurations extending in the radial direction (which is unbounded) do not exhibit regular energy/charge relations.
Instead, the criterion seems to be untrue at the quantum level on general backgrounds.

In fact, for what concerns the quantum corrections, we have generically {\it non} subleading (in $1/J$) contributions 
in the $J \rightarrow \infty$ limit, while  subleading ones for $b=2$
.
As a first conclusion, we can say that on general backgrounds the regularity of the expansion of the classical energy/charge relation do {\it not} imply that the quantum contribution is $1/J$ suppressed.
Since the latter implication is very likely to be true in the special case of $AdS_5\times S^5$, it must depend on the specific form of this background.    
But, considering the example in \cite{mina}, this property does not seem to be the huge amount of supersymmetry.
Moreover, it cannot even be the fact that the radius of the five-sphere is bounded, because we have provided a counter-example: the $b=2$ case has regular classical expansion and $1/J$ suppressed quantum corrections, but, as for the other values of $b$, the radius of the three-sphere is not fixed.
In fact, at the quadratic fluctuation level needed for the calculation of the quantum corrections, the string does see that the three-sphere radius can increase\footnote{If this would not have been the case, the $b\neq 2$ case would have provided a counter-example the other way around.}.
Note that this is visible already for the point-like strings of the pp-wave.
This interesting subject surely requires more efforts for a complete understanding.

\begin{center}
{\large  {\bf Acknowledgments}}
\end{center}
We thank F. Belgiorno, S. Cacciatori and T. Mateos for useful discussions. This work was partially supported by INFN, MURST and by the European Commission RTN program HPRN-CT-2000-00131.

\appendix
\section{The GS superstring on a general background}\label{appA}

In this appendix we review the form of the GS superstring action  expanded up to the quadratic order (see \cite{ms} and references therein). The GS type IIB superstring action expanded up to the second order in the fermions is ($\xi^\alpha,\alpha=0,1$ are world-sheet coordinates,  $x^m,m=0,\dots,9$ are the spacetime coordinates, $a,b,\ldots=0,\ldots,9$ are used for flat spacetime indexes, $\epsilon^{01}=1$):
\bea
S_{(F1)}^{IIB}&=&-\frac{1}{4\pi\alpha^\prime}\int{d^2\xi\sqrt{-h}h^{\alpha\b}\partial_\a x^m\partial_\b x^n g_{mn}}+\frac{1}{4\pi\alpha^\prime}
\int{d^2\xi\epsilon^{\a\b}\partial_\a x^m\partial_\b x^n B_{mn}} \;+\cr
&&+\;\frac{i}{4\pi\alpha^\prime}\int
d^2\xi\sqrt{-h}\;\bar{\theta}(1-\Gamma_{F1})\Gamma^\a D_\a\theta + O(\theta^4)=\cr
&=&-\frac{1}{4\pi\alpha^\prime}\int{d^2\xi\sqrt{-h}h^{\a\b}\partial_\a x^m\partial_\b x^n g_{mn}}+\frac{1}{4\pi\alpha^\prime}
\int{d^2\xi\epsilon^{\a\b}\partial_\a x^m\partial_\b x^n B_{mn}} \;+\cr
&&+\;\frac{i}{4\pi\alpha^\prime}\int
d^2\xi\big[ \sqrt{-h}h^{\a\b}\delta^{IJ}-\epsilon^{\a\b}(\sigma_3)^{IJ}\big] \bar{\theta}^I \Gamma_\a (D_j\theta)^J + O(\theta^4)\ .
\label{fIIB3}
\eea
We use the double spinor convention  
\bea
\theta=\left(
\begin{array}{cc}
\theta^1\\
\theta^2\\
\end{array}\right)\ ,
\eea
where $\theta^{1,2}$ are MW spinors in ten dimensions with positive chirality ($\Gamma^{11}\theta^{1,2}=\theta^{1,2}$). Furthermore
$\Gamma_\alpha=\partial_\alpha x^m e_m^a\Gamma_a$, the world-sheet chiral operator is 
\bea
\Gamma_{F1}=\hbox{${1\over 2\sqrt{-h}}$}\epsilon^{\a\b}\Gamma_{\a\b}\sigma_3\ ,
\eea
$D_i$ is the pull-back on the world-sheet of the operator $D_m$ defined by
\bea
\label{operatorsIIB}
D_m &=& D^{(0)}_m+W_m 
\eea
with
\begin{eqnarray}
D^{(0)}_{m} &=& \partial_m +\frac{1}{4} \omega_{mab}\Gamma^{ab}+\frac{1}{4\cdot 2!}H_{mab}\Gamma^{ab}\sigma_3\ ,\cr
W_{m} &=& \frac18 e^\Phi \left[{F}^{(1)}_a\Gamma^a(i\sigma_2) + \frac{1}{3!} {F}^{(3)}_{abc}\Gamma^{abc}\sigma_1+
\frac{1}{2\cdot 5!}{F}^{(5)}_{abcde}\Gamma^{abcde}(i\sigma_2)\right]\Gamma_m\ .
\end{eqnarray}
The operator $D_m$ enters the gravitino supersymmetry transformation
\bea\label{susyIIBfermions1}
\delta_\epsilon \psi_m=D_m \epsilon\ .
\eea
It is also useful to recall that the dilatino supersymmetry transformation takes the form
\bea\label{susyIIBfermions2}
\delta_\epsilon \lambda =\Delta\epsilon\ ,
\eea
where
\bea\label{operatorsIIB2}
\Delta &=& \Delta^{(1)}+\Delta^{(2)}\ ,\cr
&&\cr
 \Delta^{(1)} &=& \frac12 \left( \Gamma^m \partial_m\Phi +\frac{1}{2\cdot 3!}H_{abc}\Gamma^{abc}\sigma_3\right)\ ,\cr
\Delta^{(2)} &=& -\frac{1}{2} e^\Phi \left[ {F}^{(1)}_{a}\Gamma^{a}(i\sigma_2)+
\frac{1}{2\cdot 3!} { F}^{(3)}_{abc}\Gamma^{abc}\sigma_1\right]\ .
\eea
In this case the $\kappa$-symmetry reads
\bea
\delta_\kappa \theta &=& (1-\Gamma_{F1})\kappa + O(\theta^2)\ ,\cr
\delta_\kappa x^m &=& {i\over 2}\bar \theta \Gamma^m (1-\Gamma_{F1})\kappa+ O(\theta^3)\ .
\label{bkappa2}
\eea

\section{UV finiteness and stability}\label{appendixb}

In this appendix we discuss the large $n$ limit of the frequencies for the quantized circular string configuration, 
showing that the theory is UV finite, and its stability.
From the equation for the frequencies (\ref{freqs1}) one can derive the large $n$ limit of the
$\delta\chi,\,\delta\tilde{v}_1$ modes
\be
\omega_b \sim n \pm B [(m_2^2-m_1^2)+(w_1m_1-w_2m_2)]
 -\frac{a_1^2m_1^2+a_2^2m_2^2}{2n} + \frac{a_1^2w_1^2+a_2^2w_2^2}{2n}\ .
\ee
The one for the $\delta Y^a$ modes read, from (\ref{freqy}),
\bea
\omega_b&=&n \pm \frac 12 \sqrt{(M_2+M_3)^2+(M_4+M_5)^2} +\cr
&& \quad + \frac1n\Big[\frac M2+\frac{M_1}{2}-\frac{1}{8}(M_3^2-M_2^2+M_5^2-M_4^2)-\frac{M_1(M_4+M_5)^2}{4[(M_2+M_3)^2+(M_4+M_5)^2]}\Big]\ ,\cr
\omega_b&=&n+\frac1n \Big[\frac M2+\frac{M_1(M_4+M_5)^2}{2[(M_2+M_3)^2+(M_4+M_5)^2]}\Big]\ .
\eea
Taking into account the three $\delta X^a$ modes the sum of the bosonic frequencies then give
\be
\sum \omega_b = 8n + \frac{1}{n}\{(a_1^2w_1^2+a_2^2w_2^2) + \frac{b^2}{4}[2(a_1^2m_1^2+a_2^2m_2^2)+(a_1^2w_1^2+a_2^2w_2^2)]\}\ .
\ee
From the fermionic frequency equations (\ref{freqfer}) one derives the behavior
\be
\omega_f \sim n \pm (C+D) + \frac{E^2 + F^2 + G^2 + H^2}{2n}\ .
\ee
Making explicit the coefficients one can check that the sum of these eight frequencies precisely cancels the bosonic sum,
showing that the theory is indeed UV finite.\\

Finally, let us briefly discuss the stability of the theory.
As in the $AdS$ case, there exist some range of parameters where the frequencies become imaginary,
signaling an instability of the solution.
On the other hand, the important point is that this is not true in general.
Let us begin from the frequencies for the $\delta\chi,\,\delta\tilde{v}_1$ bosons, which are obviously very similar to the sphere modes in the
$AdS_5 \times S^5$ case.
As a first observation, in the single charge limit $m_2=-m_1\equiv m,\,w_2=w_1\equiv w,\,a_1=a_2=1/\sqrt{2}$, the frequency
equation (\ref{freqs1}) reduces to the analogous one in $AdS$ and so gives the same stability condition $n^2\geq 4m^2$
(cfr. (5.38) of \cite{tseyrew}).
But the interesting regime for our concern is the large $\nu^2\equiv -\Lambda$ limit (or equivalently the large $k$ limit),
where (\ref{freqs1}) has a trivially safe mode $\omega_b \sim \alpha \nu$ with real $\alpha$, and a second one going as (cfr. (5.41) of \cite{tseyrew})
\be
\omega_b \sim \frac{n}{\nu}[m_1+m_2+\frac 12 \sqrt{n^2 - 4a_1^2a_2^2(m_1-m_2)^2}]\ ,
\ee 
which gives the condition $n^2\geq 4|m_1m_2|$.

For what concerns the $\delta Y^a$ modes, one can show that there are values of the parameters where the frequency equation
(\ref{freqy}) have imaginary solutions.
But in the large $\nu$ limit one finds two trivially safe modes and a third one going as $\omega_b \sim 1/\nu$ giving a
non trivial, fourth order (in $n$),
stability condition.
A numerical analysis of the latter shows that the condition is alway satisfied, so that in the limit the $\delta Y^a$ modes give
no instabilities.

It is easy to observe from (\ref{freqfer}) that the fermionic frequencies are all safe, going as
$\omega_f \sim \beta \nu$ with real coefficients $\beta$.
In conclusion, the large $\nu$ regime has only the stability condition $n^2\geq 4|m_1m_2|$ as the two-charge circular solution
in $AdS_5 \times S^5$.
For generic $\nu$ other instabilities arise, but there is always a set of values of the other parameters for which
the theory is safe.




\end{document}